\documentclass[preprint2,longabstract]{aastex}

\shorttitle{PdBI sub-arcsecond study of the SiO microjet in HH212}
\shortauthors{Cabrit et al.}

\begin{document}


\title{PdBI sub-arcsecond study of the SiO microjet in HH212 - Origin and collimation of Class 0 jets}


\author{S. Cabrit}
\affil{LERMA, UMR 8112 du CNRS, Observatoire de Paris, 61 Av. de l'Observatoire, 75014 Paris,
France}
\email{sylvie.cabrit@obspm.fr}

\author{C. Codella}
\affil{INAF - Istituto di Radioastronomia, Sezione di Firenze, Largo E. Fermi 5, 50137
Firenze, Italy}
\email{codella@arcetri.astro.it}

\author{F. Gueth}
\affil{IRAM, 300 rue de la Piscine, 38406 Saint Martin d'H\'eres, France}
\email{gueth@iram.fr}

\author{B. Nisini}
\affil{INAF-Osservatorio Astrofisico di Roma, Via di Frascati 33,
00040 Monte Catone, Italy}
\email{nisini@oa-roma.inaf.it}

\author{A. Gusdorf}
\affil{Physics Department, The University, Durham DH1 3LE, UK}
\email{antoine.gusdorf@durham.ac.uk}

\author{C. Dougados}
\affil{Laboratoire d'Astrophysique de l'Observatoire de Grenoble, BP 53, 38041 Grenoble Cedex,
France}
\email{dougados@obs.ujf-grenoble.fr}

\and

\author{F. Bacciotti}
\affil{INAF-Osservatorio Astrofisico di Arcetri, Largo E. Fermi 5,
50125 Firenze, Italy}
\email{fran@arcetri.astro.it}




\begin{abstract}
{The bipolar HH~212 outflow has been mapped in SiO using the extended
configuration of the Plateau de Bure Interferometer (PdBI), revealing
a highly collimated SiO jet closely associated with the H$_2$ jet
component.
We study at unprecedented resolution ($0\farcs34$ across the jet
axis) the properties of the innermost SiO ``microjet'' within 1000~AU
of this young Class 0 source, to compare it with atomic microjets
from more evolved sources and to constrain its origin.
The SiO channel maps are used to investigate the microjet collimation and
velocity structure.
A large velocity gradient analysis is
applied to SiO (2--1), (5--4) and (8--7) data from the PdBI and the
Submillimeter Array to constrain the SiO opacity and abundance.
The HH212 Class 0 microjet shows striking similarities in
collimation and energetic budget with atomic microjets from T Tauri
sources. Furthermore, the SiO lines appear optically thick, unlike what is
generally assumed. We infer $T_{\rm kin}$ $\simeq $ 50--500~K
and an SiO/H$_2$ abundance $\ge 4
\times 10^{-8} - 6 \times 10^{-5}$ for $n_{\rm H_2}$ = $10^7-10^5$ cm$^{-3}$, i.e.
$0.05-90$\% of the elemental silicon.
This similar jet width, regardless of the presence of a dense envelope,
definitely rules out jet collimation by external pressure, and favors a common
MHD self-collimation (and possibly acceleration) process at all
stages of star formation. We propose that the more abundant
SiO in Class 0 jets could mainly result from rapid ($\leq$ 25 yrs)
molecular synthesis at high jet densities.}

\end{abstract}





\section{Introduction}

Millimeter interferometric studies of the L1448 and HH211 Class 0 sources
have identified high-velocity SiO jet-like emission
possibly related to the primary protostellar wind
\citep{guilloteau,chandler,hirano,palau,gueth2}. 
A similar SiO jet has
recently been discovered by Codella et al. (2007; hereafter
Paper~I) 
in the HH212 H$_2$ outflow in Orion ($d \simeq$ 450 pc) using
the new extended configuration of the PdBI. This study revealed highly
collimated SiO emission with a close spatial and kinematic
correspondence to near-IR
H$_2$ knots, indicating that both are tracing the same
molecular jet component.  In addition, an inner pair of SiO knots with
no near-IR H$_2$ counterparts was identified at $\pm$1$\farcs$5 of the
central source, with a radial velocity range pointing to a high degree
of collimation.  Continuum data at 1mm further reveal a compact,
optically thick source probably tracing a disk of diameter $\le$
120~AU. Similar conclusions were reached by Lee et al. (2007)
in a lower resolution, multi-species study with the Submillimeter
Array (SMA).

In this second paper, we further exploit the unprecedented resolution
of $0\farcs34$ HPBW across the jet axis provided by the extended
configuration of the PdBI to carry out the first comparison of the
properties of the Class 0 SiO ``microjet'' in HH212 with atomic
microjets from Class I/II sources observed on similar scales.  We
identify several similarities suggesting that the same collimation
(and possibly acceleration) mechanism is at work in Class 0 jets as in
later stages. We also demonstrate that SiO is optically thick and
close to LTE in the inner knots (as is not generally assumed) and
discuss possible origins for the abundant SiO in Class 0 jets,
compared to their more evolved counterparts.

\section{Jet collimation and velocity structure}

\subsection{Present SiO observations}

Figure~1 presents PdBI SiO 5--4 maps from Paper I
of the inner jet knots, separated into three velocity intervals. The
SiO microjet is {\it extremely narrow}, with a typical transverse
FWHM $\simeq$ $0\farcs4$ at all velocities, i.e. an intrinsic width of
$0\farcs2$ = 90~AU after correction for the PdBI HPBW of $0\farcs34$
in the same direction\footnote{A slightly larger width of $0\farcs35$
was quoted in Paper I, where cleaning had not been optimized for the
innermost jet regions.}.

Figure~1 also shows that, in both lobes, the
region of blue/red overlap is not coincident with the region of
highest radial velocities, but is slightly {\it trailing behind it} by
about $0\farcs4$. The lack of blue/red contamination towards the
fastest gas requires that its motions are highly forward-directed with
a semi-opening angle $\le 4\degr$ (see Paper I). The blue/red overlap
at low velocities $\simeq$ 3-4 km s$^{-1}$ traces less collimated,
slower material in the wake of the fastest gas.

\subsection{Comparison with jets from more evolved sources}

The width of atomic jets from T Tauri Class II sources
spans a relatively broad range, depending on the brightness of
bowshock wings driven by internal working surfaces. In
Fig.~2, the intrinsic FWHM of the HH212 SiO microjet
is compared to the broadest (DG~Tau) and narrowest (RW~Aur) atomic
microjets from Class II sources studied so far using ground-based
adaptive optics or HST \citep{dougados,woitas}.
We find that the HH212 SiO microjet falls exactly
in the same range as Class II jets on scales 500--1000 AU. Similar
results are found for the SiO jet from the HH211 Class 0 source (width
of 95--125~AU at distances of 300--600 AU; Gueth et al. 2007).

Also indicated in Fig.~2 is the width of the HH212
jet at 50 AU from the source, $\simeq$ 40 mas = 18 AU, as inferred
from the bow shape of H$_2$O maser spots within 100 mas 
\citep{claussen}. Again it is undistinguishable from that of
atomic microjets at the same distance. We thus find no evidence of a
higher jet collimation in Class 0 sources compared to the T Tauri
stage where only a thin disk is present, although the dense infalling
envelopes characterizing the Class 0 stage would be capable of
strongly reconfining a radially expanding wind 
\citep{delamarter}. This definitely rules out collimation by
external pressure gradients and requires that jets from young stellar
objects are self-collimated by internal magnetic stresses. The jet MHD
collimation process appears to be the same at all phases, with all
fast material confined within a beam diameter of about 15-20 AU over a
distance $\simeq 50$ AU.

We further note that the HH212 Class 0 microjet follows interesting
scalings compared with Class II microjets concerning its energetics
and kinematics. (i) The mass ejection to accretion rate in HH212
estimated by Lee et al. (2007) from CO emission farther out
along the jet is 15\% (scaling with $V_{\rm jet}$/100 km~s$^{-1}$). This is
similar to the ratio of 10\% found for spatially resolved Class II
jets \citep[e.g.][]{woitas}. (ii) The HH212 knot speed of
100-150 km~s$^{-1}$ is typically half that in T Tauri jets 
\citep[e.g.][]{dougados}, for a four times lower stellar mass of
0.15M$_\odot$ \citep{lee06}. Hence the jet speed
appears reduced in the same proportion as the escape speed from the
central object. Such scalings would be consistent with the jet
acceleration mechanism and launching zone also possibly being the same
at all phases. However, similar data in a larger sample of Class
0 jets would be needed to confirm this conjecture.

\section{SiO abundance in the HH~212 microjet}

\subsection{SiO line ratios and brightness temperatures}

In order to constrain the physical conditions associated with the SiO
emission in the inner jet, we compared the $J$=2--1 and 5--4 line
intensities from Paper I. For proper comparison, the SiO(5--4) map,
originally obtained with a $0\farcs78\times0\farcs34$ resolution, was
reconstructed at the lower resolution of the SiO(2--1) map
($1\farcs89\times0\farcs94$).  Figure~3 plots on a main beam
(``MB'') scale the reconstructed 5--4 line profiles at the peaks of the
inner SiO knots, and the ratio $T_{\rm MB}$ (5--4)/$T_{\rm MB}$ (2--1) as a function of
velocity (bottom panels). It can be seen that the ratio is $\simeq$
0.75--1.1 across the blue knot profile, and $\simeq$ 0.5--0.85 across
the red knot profile. Relative calibration uncertainties between the
2--1 and 5--4 lines are estimated to be $\simeq$ 20\%.  We similarly
evaluate the SiO (8--7) to (5--4) intensity ratio by degrading our
PdBI map to the $0\farcs96 \times 0\farcs69$ SMA beam of Lee et
al. (2007). The resulting (5--4) spectra towards the inner
SiO knots are also plotted in Fig.~3. Comparison with Fig.~10 of Lee
et al. (2007) yields an (8--7)/(5--4) ratio in the range
0.7--1 at all velocities. The relative calibration uncertainty could
reach 30\%.

A third constraint is provided by the peak main beam temperatures
$T_{\rm MB}$(5--4) $\simeq$ 25~K in both knots in our original PdBI beam (top curve
in Figure~3). As the jet is broadened by a factor
$\simeq$ 2 by beam convolution across the jet (cf. Sect.~2), the
intrinsic line temperature $T_{\rm R}$(5--4) is at least
25$\times$2 = 50~K. Including beam dilution along the jet axis with
$0\farcs78$ HPBW, the intrinsic line brightness could reach 200~K
if the knot is roughly circular.

\subsection{LVG modelling: evidence for optically thick SiO}

The line ratios and (5--4) intrinsic brightness are compared
with the result of a large velocity gradient (LVG) code, which
considers the first 20 levels of SiO and the rate coefficients for
collisions with H$_2$ reported by Turner et al. (1992) up to
$T_{\rm k}$ = 300~K. We explored H$_2$ densities from $10^5$ to $10^7$ cm$^{-3}$ 
(see Sect.~3.3) and an LVG optical depth parameter
$n$(SiO)/ $(dV/dz)$ = $N_{\rm SiO}/\Delta V$ ranging from $10^{12}$ to $10^{17}$ cm$^{-2}$
(km~s$^{-1}$)$^{-1}$, i.e. from the fully optically thin to optically thick
regime. Our typical model results are illustrated graphically for
$T_{\rm k}$ = 100~K in Fig.~4, and compared with observed
values in HH212.

We find that the usual approach of assuming optically thin emission to
derive $n({\rm H_2})$ and $T_{\rm k}$ from line ratios \citep[e.g.][]{gibb,nisini}
would give inaccurate results in our
case: As shown in Fig.~4(bottom panel), no optically thin model
(starred symbols in the curves) can {\it simultaneously} reproduce the
observed values of both SiO(8--7)/(5--4) and SiO(5--4)/(2--1).  Values
$\simeq$ 1 for both ratios are only achieved when approaching the
optically thick LTE regime ($T_{\rm R} \simeq$ $T_{\rm k}$),
which is the point of convergence of all density curves at
sufficiently high opacity. We infer that
$N_{\rm SiO}/\Delta V$ must be greater than $\simeq 10^{15}$ cm$^{-2}$ 
(km~s$^{-1}$)$^{-1}$, while
$n({\rm H_2})$ is not well-constrained. The high (5--4)
intrinsic brightness of 50~K-200~K also independently argues for a
large optical depth parameter (Fig.~4, top). It also
indicates that $T_{\rm k}$ lies in the range 50-500 K, or else
the predicted $T_{\rm R}$(5--4) close to LTE would be too low/high.

We note that substantial SiO optical depth could be rather common in
the innermost part of Class 0 jets, if they are as narrow as in
HH212. In the L1448 jet, for example, a column density $\sim$10$^{14}$
cm$^{-2}$ has been derived from single-dish measurements of the 5--4
transition assuming a jet width of 2$\arcsec$ 
\citep{nisini}. A narrower width of $\sim$0.2$\arcsec$ would
result in a column density higher by an order of magnitude, implying,
as in HH212, a line optical depth larger than unity. Hence the low
$T_{\rm MB}$(5--4) $\simeq$ 0.1-1~K in single-dish observations could result mainly
from severe beam dilution of the SiO emission, as argued
previously by Gibb et al. (2004) and illustrated in Fig.~{
3}. The SiO abundances would then be substantially larger than previously
reported.

\subsection{SiO abundance and H$_2$ density}

Noting that $N_{\rm SiO}/\Delta V$ = $n$(SiO)/ $(dV/dz)$, the SiO abundance with
respect to H$_2$ may be written:
\small{
\begin{equation}
X({\rm SiO}) = 4\times10^{-7}
\left({N_{\rm SiO}/\Delta V \over 10^{15} {\rm cm^{-2} km^{-1} s}}\right)
\left({10^6 \over n(H_2)}\right) 
\left({ dV/dz \over 4\,10^{-11} {\rm s^{-1}}}\right).
\end{equation}}
The adopted line-of-sight velocity gradient $dV/dz$ is typical of
cooled regions with $T_{\rm k}$ $\le$ 100 K at the rear of planar
C-shocks and is probably a lower limit. A steeper gradient $dV/dz$ is
given by the ratio of the FWZI of the SiO line profile ($\simeq$ 10
km~s$^{-1}$) to the knot width  ($0\farcs2$ = 100~AU), which
would increase X(SiO) by a factor 16 from the above formula.

The main uncertainty in X(SiO) stems from the unknown H$_2$ density in
the SiO knots.  A reasonable range may be inferred from the presence
of shock-excited H$_2$O masers at 0.1$\arcsec$ from the
source. Magnetic field strengths and line ratios in H$_2$O masers
around YSOs typically require preshock H nuclei densities $n_{\rm H}$ $\simeq
10^7-10^8$ cm$^{-3}$ \citep{KN}. Assuming that density
roughly drops with distance as $1/r^2$ (cf the DG Tau jet;
Lavalley-Fouquet et al. 2000), one infers a preshock
density $\simeq 10^5-10^6$ cm$^{-3}$ at the SiO knots. Shock compression
could increase these values by a about an order or magnitude
\citep[e.g.][]{KN}, so that the density is in the range
$\simeq 10^5-10^7$ cm$^{-3}$. The resulting {\it minimum} SiO abundance for
optically thick emission is X(SiO) $\ge (4 \times 10^{-8} - 4 \times
10^{-6})\times(1-16)$, with the higher value corresponding to the
lower density, and the additional factor 1--16 arising from the
uncertainty in velocity gradient. Assuming a solar abundance of
(Si/H)$_\odot$ $\simeq 3.5 \times 10^{-5}$
\citep{si-abun}, between 0.05\% and 90\% of the elemental silicon
is in the form of SiO.

\subsection{Origin of the SiO component}

Our PdBI observations of the HH212 microjet set stronger constraints than
previously on the origin of the SiO in protostellar outflows,
because of the shorter timescales involved and the unusually high
collimation and SiO column densities indicated by our data.

Given the proper motions of 60-150 km~s$^{-1}$ for H$_2$O masers and
H$_2$ knots \citep{claussen,mc02}, 
the dynamical time of inner SiO peaks at 500~AU is
only 25 yrs. SiO should thus be incorporated very rapidly in the
flow.  The formation of SiO in outflows is usually attributed to
sputtering of Si atoms from charged grains in a magnetized C-shock
with ion-neutral drift speeds $\ge$ 25 km~s$^{-1}$ 
\citep{schilke}. Updated C-shock models with improved sputtering
yields, SiO formation rates, and molecular cooling (Gusdorf et al., in
preparation) show that the required conditions for optically thick
emission are reached for shock speeds 35--45 km~s$^{-1}$ and preshock
densities of $10^5-10^6$ cm$^{-3}$ but only at the rear of the shock
where velocity gradients are small, i.e.  after 400--150 yrs. As this
exceeds the knot dynamical time, non-steady truncated C-shocks need to
be considered to model SiO-emitting shocks on such small scales.

Another long-standing issue is whether SiO molecules originate from
shocked ambient material or trace the primary jet itself. The SiO
microjet diameter of 100 AU is comparable to the centrifugal disk
diameter of 120~AU indicated by our 1mm continuum size (Paper~I) and
by envelope kinematics \citep{lee06}. Hence we would
expect little infalling molecular material left on-axis to refill the
jet path between successive ejection episodes, unless this material is
very warm. This would appear to favor an origin of the SiO in the jet
itself. The option is appealing, as the higher densities of Class 0
jets, and the accompanying low temperature and ionization, are indeed
conducive to molecular formation. In an early study of chemistry in
protostellar winds, Glassgold et al. (1991) found that Si
atoms are quickly converted into SiO at high mass-loss rates ${\dot M}_{\rm jet}$~ $>
10^{-6}$M$_\odot$~yr$^{-1}$. For a dust-free wind, the predicted SiO abundances
are $\simeq$ 50-100\% of the total elemental silicon. However, the
recent finding of a substantial depletion of Fe and Ca at the base of
several Class I jets \citep{podio} indicates that jets
are not dust-free and that grains are only partly eroded along the
flow. In the HH34 jet, 13\% of Fe has been returned to the gas at
distances $\ge$ 1500 AU. The same process at work in Class 0 jets
would release Si atoms in a sufficient amount to produce optically thick
SiO emission if ${\dot M}_{\rm jet}$ $\ge 10^{-6}$M$_\odot$~yr$^{-1}$. In the inner SiO knots of
HH212, this mass-flux is achieved for $n({\rm H_2})$ $\ge 10^6$cm$^{-3}$ (with
$V_{\rm jet} = $ 100 km~s$^{-1}$ and a jet radius of 50~AU), thus only
0.5\%--9\% of Si would be needed, if all is converted into SiO (see
Eq.~1).

\section{Conclusions}

Our finding that jet collimation in the HH212 Class 0 source is
similar to that in T Tauri stars favors a collimation mechanism
independent of the presence of a dense envelope, i.e. most probably
internal MHD stresses. The ejection/accretion ratio and the jet
speed/escape speed ratio also appear to be similar to those in Class
II, possibly suggesting the same acceleration mechanism as well.  The
main difference between Class 0 jets and their more evolved analogs
would then be their differing chemical composition, with
abundant molecules at the Class 0 stage, a mixed atomic-molecular
composition at the Class I stage \citep{davisH2,davisFeII},
and a purely atomic flow at the Class II stage.

We also find that SiO is optically thick, so that its abundance
is larger than previously estimated. The extremely narrow width of the SiO
jet revealed by PdBI further argues that this species is not formed in
swept-up material, but more likely within the jet itself. We thus
propose that the higher SiO content of Class 0 jets could mainly reflect
an increase in jet density (hence, a higher efficiency of
molecular formation), linked to the increased mass-accretion rate at
earlier stages.

\acknowledgments

We are grateful to R. Cesaroni, J. Ferreira, and an anonymous referee
for helpful comments.  This work is supported in part by the
European Community's Marie Curie Research Training Network JETSET
under contract MRTN-CT-2004-005592. It benefited from research funding
by the European Community's sixth Framework Programme under RadioNet
R113CT 2003 5058187. A. Gusdorf acknowledges support through the
European Community's Human Potential Programme under contract
MRTN-CT-2004-512302, Molecular Universe.

\clearpage



\begin{figure*}
\includegraphics[angle=-90,scale=0.75]{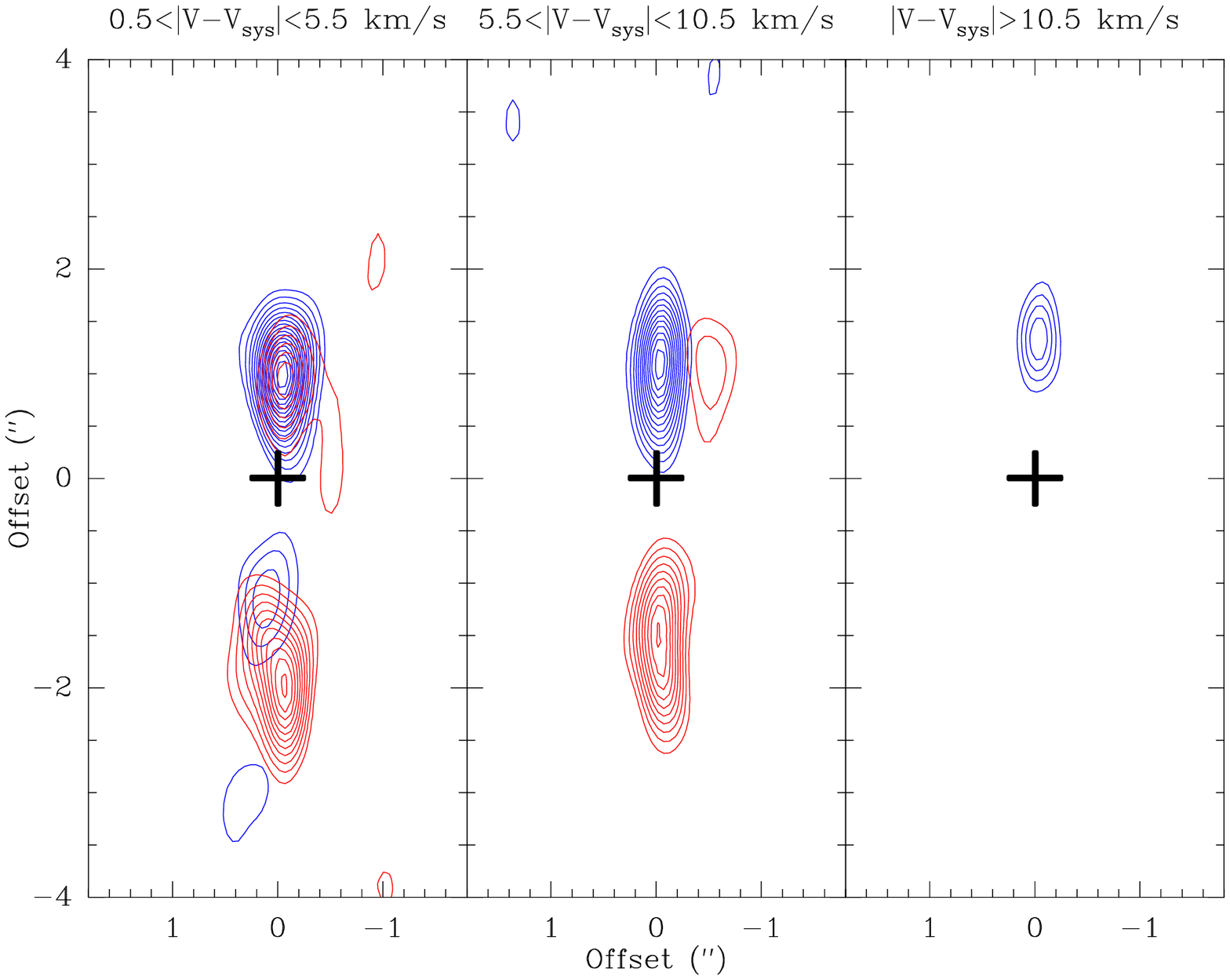}
\caption{SiO (5--4) emission maps of the HH212 microjet in 3
different velocity ranges. Blue and red contours refer to
blueshifted and redshifted gas, respectively.
A cross marks the position of the continuum source from Paper I:
$\alpha$(2000) = 05$^h$ 43$^m$ 51$^s$.41, $\delta$(2000) =
-01$\degr$ 02$\arcmin$ 53$\farcs$160.
Contour spacing is 50 mJy/beam km~s$^{-1}$ with the first contour at 100
mJy/beam km~s$^{-1}$.}
\end{figure*}

\begin{figure*}
\includegraphics[angle=-90,scale=0.6]{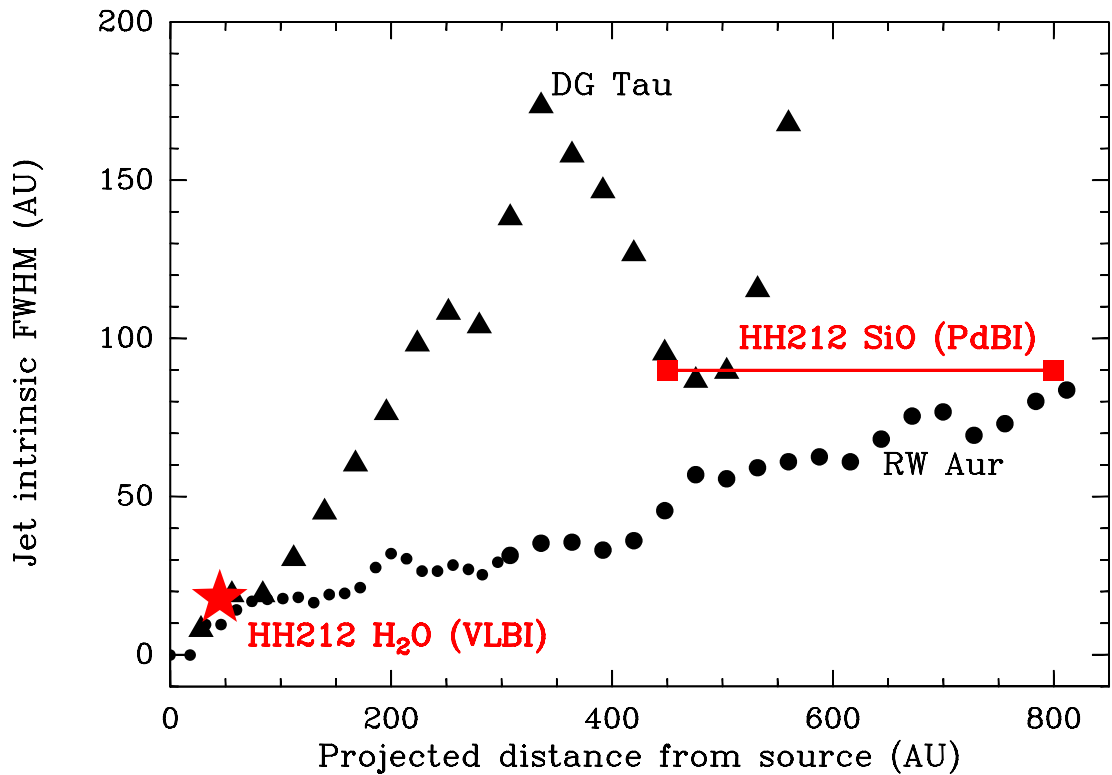}
\caption{HH212 intrinsic jet width compared to the range spanned by
atomic microjets from Class II sources, corrected for the
instrumental PSF (small dots: Woitas et al. 2002; large dots and triangles:
Dougados et al. 2000); Our SiO PdBI measurements are shown as filled
squares; the H$_2$O maser width from Claussen et al. (1998) as a filled
star.}
\end{figure*}

\begin{figure*}
\includegraphics[angle=-90,scale=0.6]{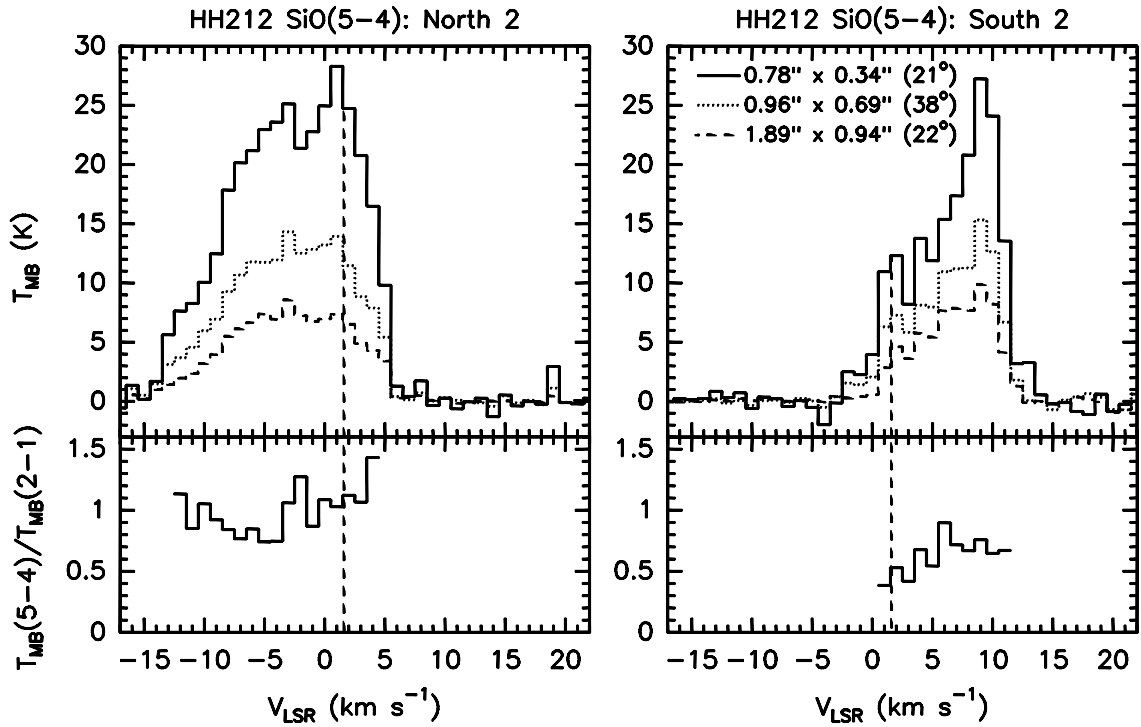}
\caption{Top panels: Line profiles in SiO $J$=5--4 towards the inner
SiO knots at various resolutions: the original PdBI beam (solid
histogram), the SMA $J$=8--7 beam (dotted histogram), and the PdBI SiO
$J$=2--1 beam (dashed histogram). Beam PAs are listed between
parentheses. Note the dramatic decrease in brightness temperature
with increasing beam dilution. The vertical dashed line marks the
ambient LSR velocity (+1.6 km s$^{-1}$; Wiseman et al. 2001). Bottom
panels: Line temperature ratio $T_{\rm MB}$(5--4)/$T_{\rm MB}$(2--1)
at the resolution of the PdBI SiO $J$=2--1 map, as a
function of velocity.}
\end{figure*}

\begin{figure*}
\includegraphics[angle=-90,scale=0.6]{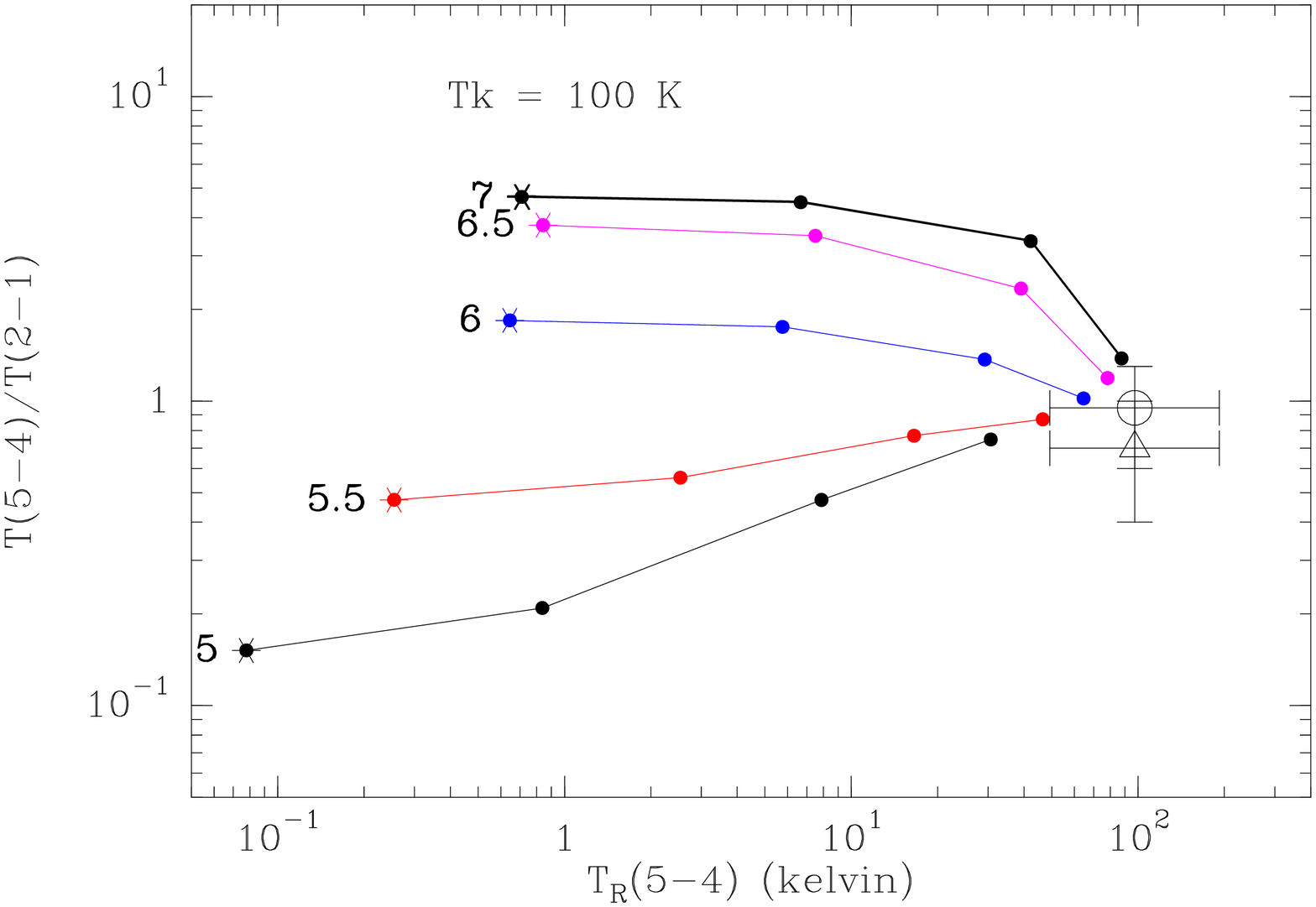}
\includegraphics[angle=-90,scale=0.6]{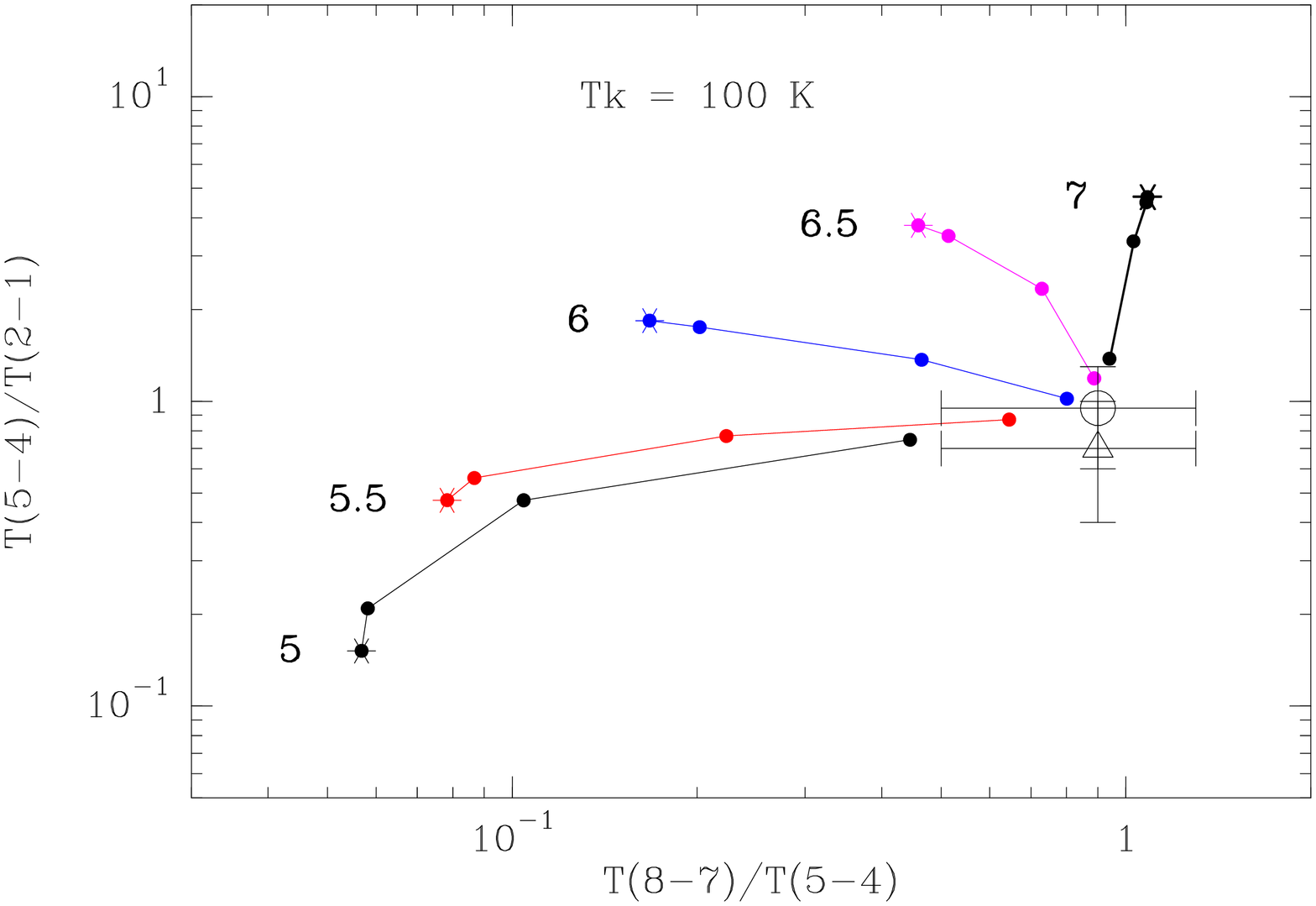}
\caption{Top: SiO line temperature
ratio $T$(5--4)/$T$(2--1) versus intrinsic line temperature $T_{\rm R}$(5--4) for
LVG slab models at $T_{\rm k}$ = 100~K. Each curve corresponds to the
labelled log($n({\rm H_2})$), with dots marking values of
$N_{\rm SiO}/\Delta V$ increasing (left to right) from $10^{12}$ to $10^{15}$ cm$^{-2}$ 
(km~s$^{-1}$)$^{-1}$ by factors of 10. Symbols with error bars illustrate the
range in line ratio in the inner SiO knots of HH212 (including
calibration uncertainties) and the range in $T_{\rm R}$(5--4) 
after correction for beam dilution. Bottom: Same as above for the
$T$(5--4)/$T$(2--1) ratio versus $T$(8--7)/$T$(5--4).}
\end{figure*}

\end{document}